\documentclass[sigconf,nonacm]{acmart}
\usepackage{xspace}
\newcommand{\harmonica}{\protect{$\mathcal{H}armonica$}\xspace}
\newcommand{\harmone}{\protect{$\mathcal{H}armon{E}$}\xspace}
\AtBeginDocument{%
  }
\setcopyright{acmlicensed}
\copyrightyear{2025}
\acmYear{2025}
\acmDOI{XXXXXXX.XXXXXXX}
\acmConference[ICML-Workshop '25]{Workshop on Sustainable MLOps}{2025}{Somewhere}
\acmISBN{978-1-4503-XXXX-X/2025/07}
\usepackage{float}  

\begin{document}

\title{\harmonica: A Self-Adaptation Exemplar for Sustainable MLOps}

\author{Ananya Halgatti}
\orcid{0009-0005-6322-337X}
\affiliation{%
  \institution{SERC, IIIT-Hyderabad}
  \city{Hyderabad}
  \country{India}}
\email{ananya.halgatti@students.iiit.ac.in}

\author{Shaunak Biswas}
\orcid{0009-0009-7207-0829}
\affiliation{%
  \institution{SERC, IIIT-Hyderabad}
  \city{Hyderabad}
  \country{India}}
\email{shaunak.biswas@research.iiit.ac.in}

\author{Hiya Bhatt}
\orcid{0009-0000-1773-5327}
\affiliation{%
  \institution{SERC, IIIT-Hyderabad}
  \city{Hyderabad}
  \country{India}}
\email{hiya.bhatt@research.iiit.ac.in}

\author{Srinivasan Rakhunathan}
\orcid{0009-0003-6507-9485}
\affiliation{%
  \institution{Microsoft, SERC, IIIT-Hyderabad}
  \city{Hyderabad}
  \country{India}}
\email{srinivasan.r@research.iiit.ac.in}

\author{Karthik Vaidhyanathan}
\orcid{0000-0003-2317-6175}
\affiliation{%
  \institution{SERC, IIIT-Hyderabad}
  \city{Hyderabad}
  \country{India}}
\email{karthik.vaidhyanathan@iiit.ac.in}

\renewcommand{\shortauthors}{}

\begin{abstract}
Machine learning enabled systems (MLS) often operate in settings where they regularly encounter uncertainties arising from changes in their surrounding environment. Without structured oversight, such changes can degrade model behavior, increase operational cost, and reduce the usefulness of deployed systems. Although Machine Learning Operations (MLOps) streamlines the lifecycle of ML models, it provides limited support for addressing runtime uncertainties that influence the longer term sustainability of MLS. To support continued viability, these systems need a mechanism that detects when execution drifts outside acceptable bounds and adjusts system behavior in response. Despite the growing interest in sustainable and self-adaptive MLS, there has been limited work towards exemplars that allow researchers to study these challenges in MLOps pipelines. This paper presents \harmonica, a self-adaptation exemplar built on the \harmone approach, designed to enable the sustainable operation of such pipelines. \harmonica introduces structured adaptive control through MAPE-K loop, separating high-level adaptation policy from low-level tactic execution. It continuously monitors sustainability metrics, evaluates them against dynamic adaptation boundaries, and automatically triggers architectural tactics when thresholds are violated. We demonstrate the tool through case studies in time series regression and computer vision, examining its ability to improve system stability and reduce manual intervention. The results show that \harmonica offers a practical and reusable foundation for enabling adaptive behavior in MLS that rely on MLOps pipelines for sustained operation.

\end{abstract}

\ccsdesc[500]{Software and its engineering~Designing software}
\ccsdesc[500]{Social and professional topics~Sustainability}
\ccsdesc[500]{Software and its engineering~Software architectures}
\ccsdesc[500]{Computing methodologies~Machine learning}
\ccsdesc[500]{Social and professional topics~Software selection and adaptation}
\keywords{Self-Adaptation, MLOps, Exemplar, Sustainability, Green AI}

\maketitle


\section{Introduction}

Machine Learning Enabled Systems (MLS) are increasingly deployed in real-world settings where they must operate under dynamic and often unpredictable conditions \cite{tamburri2020sustainablemlops}. These shifting environments introduce runtime uncertainties that affect the behaviour of the ML components embedded in such systems. Although Machine Learning Operations (MLOps) provides structured processes for developing, deploying, and monitoring ML models, its practices largely emphasise reliability and automation, which support technical sustainability by helping systems maintain their functionality over time. However, these practices do not address sustainability in a broader sense, nor do they provide systematic runtime adaptation mechanisms that account for factors such as energy usage or operational cost. Sustainability extends beyond technical performance to include environmental, economic, and social dimensions as well~\cite{karlskrona,lago_sus_dimension}, all of which shape the long-term viability of MLS.

Recognising that sustainable operation of MLS requires managing trade-offs across these dimensions has motivated the application of self-adaptation \cite{danny_sa, maria_thesis}, where feedback loops are used to react to performance degradation or energy fluctuations during execution~\cite{tedla2024ecomls,edgemlakhila}. Architectural work such as \harmone~\cite{harmone_hiya} further demonstrates how sustainability metrics including predictive accuracy, energy consumption, and distribution shifts can guide adaptation at runtime when bounded by design-time sustainability goals expressed through adaptation boundaries~\cite{adaptation_as_sus_goal_ilias}. While these approaches show how sustainability-aware decisions can be embedded into MLS, they are implemented as tightly coupled case study prototypes, making it difficult for the broader community to apply and extend them.

A key challenge is the absence of reusable exemplars that operationalise sustainability-aware self-adaptation in MLOps pipelines. Without such tools, researchers must reimplement monitoring pipelines and adaptation logic, which raises the barrier to entry and limits both reproducibility and systematic exploration of alternative adaptation strategies. Existing MLS exemplars focus on inference-time model switching \cite{marda2024switch} but do not support sustainability goals, retraining workflows, or lifecycle-wide behaviour, leaving a gap for experimentation with end-to-end MLOps pipelines.

This paper addresses this gap by presenting \harmonica \footnote{\href{https://github.com/sa4s-serc/HarmonE-tool}{https://github.com/sa4s-serc/HarmonE-tool}\\ \href{https://youtu.be/oiH0wzCPKSs}{https://youtu.be/oiH0wzCPKSs}}, a self-adaptation exemplar built on the \harmone architectural approach to support sustainability-aware adaptation in MLOps pipelines. \harmonica can be instantiated and configured for different MLS scenarios and provides a managing system that integrates a configurable MAPE-K loop~\cite{kephart2003vision} with existing pipelines without requiring changes to the training or inference setup of the managed system. It supports the specification of adaptation intent through sustainability goals and adaptation boundaries at design time, and at runtime it collects and analyses sustainability metrics while executing adaptation decisions. The main contributions of this work are: \textit{(i)} providing a configurable managing system that executes adaptation logic over any compatible pipeline \textit{(ii)} enabling users to upload their own datasets to \harmonica through its interface, \textit{(iii)} offering built-in visualisation that presents live sustainability and performance metrics through interactive dashboards, and \textit{(iv)} supporting offline analysis through downloadable telemetry and log data. Through these capabilities, \harmonica offers a practical and reusable platform for inspecting system behaviour, experimenting with sustainability-aware adaptation strategies, and conducting repeatable empirical studies.

\begin{figure*}[h!]
  \centering
  \includegraphics[width=0.9\textwidth]{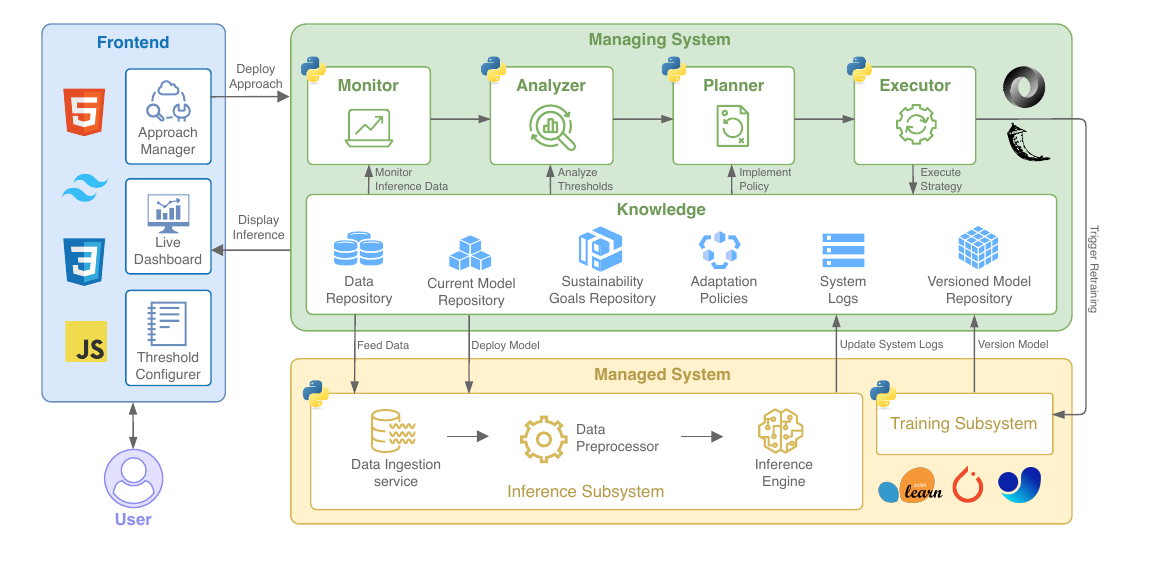}
  \caption{\harmonica architecture showing main components and data flow}
  \Description{Block diagram of the \harmonica tool showing the interaction between the Frontend, Managing System, and Managed System.}
  \label{fig:harmonica}
\end{figure*}
We evaluate \harmonica through a traffic prediction use case implemented in two domains: time series regression and computer vision. In each case, \harmonica is integrated into existing MLOps pipelines to study how the exemplar handles sustainability oriented trade offs and adapts under runtime uncertainty. We further validated the exemplar through a user study, which showed that users could configure adaptation goals with ease and follow the system’s behavior during execution while comparing alternative adaptation strategies in a controlled setting.

\section{Related Work}

Self-adaptation \cite{danny_sa} has been applied to MLS to manage runtime uncertainty \cite{maria_thesis}. Approaches such as AdaMLS~\cite{kulkarni2023towards} and energy-aware techniques like EcoMLS~\cite{tedla2024ecomls} adjust system behaviour to react to changing workloads or energy conditions. While these efforts demonstrate the value of runtime decision-making, they operate primarily at the inference layer and do not address the sustainability consequences arising across the full ML lifecycle, where retraining and model evolution introduce significant computational costs \cite{strubell}. In parallel, MLOps research has advanced automation for training, deployment, and monitoring~\cite{symeonidis2022mlops,amershi2019software}. However, these frameworks typically treat sustainability as a secondary optimisation outcome rather than an explicit architectural requirement~\cite{lago_sus_dimension}. To make sustainability explicit, architectural work has proposed design-time sustainability goals classification via decision maps~\cite{adaptation_as_sus_goal_ilias} that express adaptation intent through thresholds on metrics such as accuracy or energy consumption. These contributions show how sustainability goals can guide runtime decisions through the MAPE-K loop \cite{kephart2003vision}, though design-oriented tools like the SAF Toolkit~\cite{saf2025} currently lack the runtime mechanisms to enforce such goals in continuous ML workflows. Despite these advances, the community lacks reusable exemplars to validate sustainability-aware adaptation in real world settings. Most of the existing self-adaptive exemplars \cite{swim2018, deltaiot, SelfAdaptiveExemplars} focus on service-based systems and do not incorporate ML components. Even MLS-oriented exemplars ~\cite{marda2024switch} provide only inference-level adaptation and do not expose sustainability goals or retraining workflows. To the best of our knowledge, no existing exemplar supports adaptive control across the full ML lifecycle while providing visibility into sustainability intent. \harmonica addresses this gap by integrating a configurable MAPE-K loop into MLOps pipelines, supporting both inference and retraining, and exposing sustainability metrics in a reproducible experimental environment.

\section{Architecture and Design}
\label{sec:architecture}

As shown in Figure \ref{fig:harmonica}, \harmonica consists of three components. The \textbf{Frontend} enables user interaction, the \textbf{Managing System} performs adaptation through a MAPE K loop, and the \textbf{Managed System} provides the ML pipeline being adapted.

\vspace{1mm}
\noindent\textbf{ \large 3.1 Frontend}

\noindent The Frontend provides the interaction layer for configuring sustainability goals and managing the execution of adaptation strategies. It is implemented as a lightweight web application using \texttt{HTML, CSS (Tailwind)}, and \texttt{JavaScript}, with runtime telemetry delivered from the \texttt{Flask} backend for low latency visualization. As shown in Figure \ref{fig:harmonica}, it contains three primary components:

\noindent\textbf{\textit{3.1.1 Approach Manager}} allows users to select an adaptation approach. It supports predefined adaptation logic including: \textit{i)} model switching, \textit{ii) \harmone approach}, as well as \textit{iii)} custom strategies defined by the user. Once an approach is selected, it is deployed to the \textit{Managing System} through a \texttt{REST} endpoints exposed by the \texttt{Flask API} which load \texttt{JSON}-based policy files into the backend’s \textit{Knowledge}. The selected approach defines how monitoring data will be interpreted and how boundary violations will trigger adaptation.

\noindent\textbf{\textit{3.1.2 Threshold Configurer}} gives users control over the sustainability goals by defining adaptation boundaries at design time. These goals define the acceptable operating region of the system. The interface allows users to specify upper or lower thresholds for metrics such as energy consumption, inference accuracy, or cost related indicators. These configured thresholds are transmitted to the backend via \texttt{REST} and stored in the \textit{Sustainability Goals Repository} inside \textit{Knowledge}.

\begin{figure}[h]
  \centering
  \includegraphics[width=0.484\textwidth]{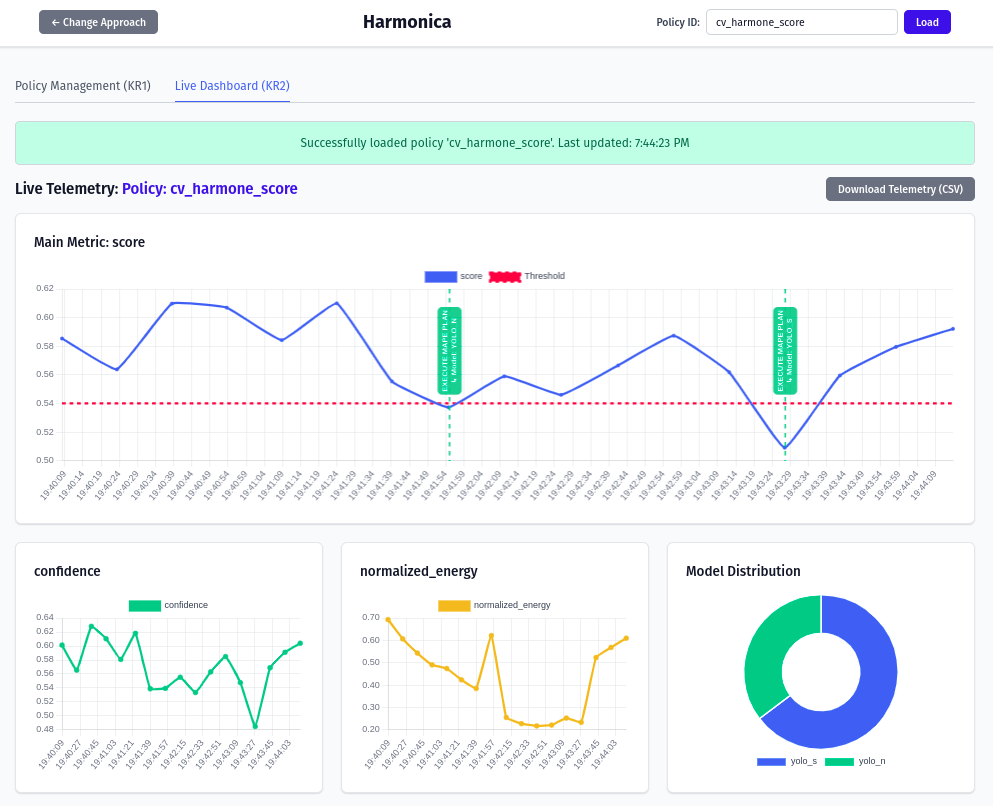}
  \caption{\harmonica dashboard}
  \Description{Block diagram of the \harmonica tool showing the interaction between the Frontend, Managing System, and Managed System.}
  \label{fig:dashboard}
\end{figure}

\noindent\textbf{\textit{3.1.3 Live Dashboard}} provides runtime observability. As shown in Figure~\ref{fig:dashboard}, the dashboard visualizes the monitored metrics in relation to the configured thresholds and displays adaptation events. It is designed as an interactive visualization that supports experimentation with adaptation strategies. The dashboard receives inference results from \textit{System Logs} from \textit{Knowledge} in real time, making it possible to analyze the effect of different configurations and adaptation approaches.

\vspace{1mm}
\noindent\textbf{\large 3.2 Managing System}

\noindent The Managing System, implemented as a \texttt{Python} \texttt{Flask} server drives self adaptation using the MAPE K loop and manages the knowledge required for runtime decision making. It receives sustainability goals and approach definitions from the Frontend and orchestrates the execution of tactics on the Managed System. 

\noindent\textbf{\textit{3.2.1 Knowledge}} consists of the following components, as illustrated in Figure~\ref{fig:harmonica}: \textit{i) Data Repository:} stores incoming inference data required for computing performance metrics and drift indicators \cite{augur}. \textit{ii) System Logs:} maintains historical logs of inference results and adaptation events in \texttt{CSV} files. Logs provide transparency and support later analysis. \textit{iii) Sustainability Goals Repository:} stores policy files in \texttt{JSON} format with the thresholds configured by the user through the \textit{Threshold Configurer}. \textit{iv) Current Model Repository:} contains the spectrum of ML models which vary in accuracy and energy use, allowing the system to tune trade offs between multiple sustainability dimensions. While \harmonica provides built-in model integration, users may upload their own models or configuration files to extend this repository and evaluate custom adaptation scenarios. \textit{v) Adaptation Policies:} includes predefined adaptation logic in \texttt{JSON} configuration, aligned with the \harmone architecture, switching based tactics, as well as custom policies. \textit{vi) Versioned Model Repository:} stores previous versions of models and their associated data received from the \textit{Training Subsystem}. This enables reuse of suitable model versions when similar conditions recur. 

\noindent\textbf{\textit{3.2.2 MAPE}}, as shown in Figure \ref{fig:harmonica}, consists of four components operating over shared \textit{Knowledge} implemented in \texttt{Python}. The \textit{Monitor} collects runtime metrics such as accuracy, confidence, and energy consumption from the \textit{System logs} in the \textit{Knowledge}. The \textit{Analyzer} interprets these metrics against the adaptation boundaries defined by the user at design time and stored in the \textit{Sustainability Goals Repository} inside \textit{Knowledge}, checking for drift outside acceptable bounds. When a violation is detected, the \textit{Planner} determines an adaptation tactic based on the \textit{Adaptation Policies} stored in the \textit{Knowledge} associated with the deployed approach. The \textit{Executor} carries out the selected tactic on the \textit{Managed System} and updates the \textit{System Logs}.

\vspace{1mm}
\noindent\textbf{{\large 3.3 Managed System}}

\noindent The Managed System, implemented in \texttt{Python} contains the ML pipeline that is subject to adaptation. It consists of the following components:

\noindent\textbf{\textit{3.3.1 Inference Subsystem}}
executes inference requests using the model currently deployed by the \textit{Managing System}. As shown in Figure~\ref{fig:harmonica}, the subsystem includes: \textit{i) Data Ingestion Service:} receives the incoming data stream from the \textit{Data Repository} in \textit{Knowledge}, representing operational inputs that may exhibit distribution shifts. The current implementation reads data from local files, but the service can be extended to ingest live streams as well.
 \textit{ii) Data Preprocessor:} prepares incoming data for inference. It extracts features or structures data into the expected representation. \textit{iii) Inference Engine:} uses the deployed model for making inferences. Models are loaded from the \textit{Current Model Repository} inside \textit{Knowledge} based on the command received from the \textit{Executor}. These inferences are then stored in \textit{System Logs} inside \textit{Knowledge}.

\noindent\textbf{\textit{3.3.2 Training Subsystem}} is triggered by the \textit{Executor} when \textit{Planner} selects retraining as a tactic. It retrains or fine tunes the model, produces updated model weights, and stores the updated model and associated training data in the \textit{Versioned Model Repository}. This allows the \textit{Managing System} to reuse suitable models and avoid redundant retraining.

\section{Empirical Evaluation}
\label{sec:empirical_eval}
This section evaluates the utility of \harmonica as an exemplar for sustainability-aware self-adaptation in ML pipelines. The goal is to demonstrate how the tool enables controlled experimentation with adaptive strategies, rather than to introduce new adaptation methods. To this end, we deploy the various adaptation scenarios within \harmonica across the following representative MLS settings: a time series regression task for traffic flow prediction and a computer vision task for object detection using diverse real-world driving scenes. In both settings, \harmonica provides the instrumentation required to observe runtime behavior, log adaptation events, and assess trade-offs between predictive quality, latency, and energy consumption.

\noindent\textbf{System Specifications: }All experiments are conducted on Pop!\_OS 22.04 LTS (x86\_64) with a 12th Gen Intel Core i5-1235U processor and 16 GB RAM. Energy consumption is measured using \texttt{pyRAPL}. The sustainability goals for \harmone are encoded as dynamic adaptation boundaries stored in \harmonica's \textit{Sustainability Goals Repository} inside \textit{Knowledge} (as shown in Figure \ref{fig:harmonica}). These goals are fully user configurable and can be defined either through \harmonica’s \textit{Threshold Configurer} interface or by supplying custom configuration files, which the tool automatically integrates into the adaptation process.

\vspace{2mm}
\noindent\textbf{\large 4.1 Use Case 1: Time Series Regression}

\noindent\textbf{\textit{Context and Setup}}
We use traffic flow data (\texttt{Vehicles/5 Minutes}) from the California PeMS\footnote{\href{https://pems.dot.ca.gov/}{https://pems.dot.ca.gov}} platform to predict short term flow values. The \textit{Current Model Repository} in \textit{Knowledge} includes three models: \texttt{scikit-learn} implementations of \textit{Linear Regression (LR)} and \textit{Support Vector Machine (SVM)}, and a \texttt{PyTorch}-based \textit{Long Short Term Memory (LSTM)}, with the order \textit{LR < SVM < LSTM} in terms of both accuracy and energy usage. Each model receives the last five timesteps as input to predict the flow at the sixth timestep. Drift is induced using a controlled scale and shift transformation applied to two chunks of the evaluation data to evaluate drift detection and model reuse behavior.

\noindent\textbf{\textit{Baselines and Results}}
We compare static baselines, a naive switching heuristic that alternates among LR, SVM, and LSTM without retraining, and \harmone running within \harmonica. Table~\ref{tab:ts_results} summarizes the results.

\begin{table}[h]
\centering
\begin{tabular}{lccc}
\hline
\textbf{Approach} & \textbf{$R^2$} & \textbf{Time (s)} & \textbf{Energy (J)} \\
\hline
LR & 0.7244 & 0.000437 & 8344.59 \\
SVM & 0.8205 & 0.000475 & 9090.09 \\
LSTM & 0.9171 & 0.003258 & 76729.03 \\
Switch & 0.8628 & 0.001501 & 28606.79 \\
\harmone & 0.8938 & 0.002433 & 42645.82 \\
\hline
\end{tabular}
\caption{Time series regression results in \harmonica.}
\label{tab:ts_results}
\end{table}

\noindent\textbf{\textit{Analysis}}
Static LSTM yields the highest accuracy but incurs more than ten times the energy cost of SVM and LR. Simple switching reduces energy but suffers from higher latency and lower accuracy than \harmone. As expected, running \harmone within \harmonica provides a balance by adapting MLOps pipelines to runtime uncertainties. The dynamic boundary mechanism reduces unnecessary interventions and avoids over adaptation.

\vspace{1mm}
\noindent\textbf{\large 4.2 Use Case 2: Object Detection}

\noindent\textbf{\textit{Context and Setup}}
For the second use case, we evaluate \harmonica on a computer vision task using the BDD100K dataset\footnote{\href{http://bdd-data.berkeley.edu/}{http://bdd-data.berkeley.edu/}}, which contains diverse real world driving scenes. The goal is to detect objects such as vehicles and pedestrians under dynamic data characteristics reflecting operational uncertainty. The \textit{Current Model Repository} in \textit{Knowledge} includes three \texttt{Ultralytics} \textit{YOLOv8} variants: \textit{YOLOv8n}, \textit{YOLOv8s}, and \textit{YOLOv8m}, ordered by increasing accuracy and energy usage \textit{(YOLOv8n < YOLOv8s < YOLOv8m)}. These models are provided as built in examples to demonstrate \harmonica's capabilities.

\noindent\textbf{\textit{Baselines and Results}}
We compare single model baselines using a fixed YOLOv8 model and a switching heuristic that alternates among the three variants executed within \harmonica. Table~\ref{tab:cv_results} summarizes the results.

\begin{table}[h]
\centering
\begin{tabular}{lccc}
\hline
\textbf{Approach} & \textbf{Confidence} & \textbf{Time (s)} & \textbf{Energy (J)} \\
\hline
YOLOv8n & 0.5163 & 0.1828 & 2.954 \\
YOLOv8s & 0.5276 & 0.3717 & 6.130 \\
YOLOv8m & 0.5679 & 0.7831 & 12.634 \\
Switch & 0.5262 & 0.3166 & 5.153 \\
\hline
\end{tabular}
\caption{Computer Vision results in \harmonica}
\label{tab:cv_results}
\end{table}

\noindent\textbf{\textit{Analysis}}
The results for the computer vision scenario demonstrate that \harmonica can support experimentation with object detection in addition to time series pipelines. Since the original \harmone approach was not generalised for this domain, it was not applied here. We believe that with domain specific MAPE-K files and customised tactic definitions, the approach could be extended to computer vision tasks. The results shown in Table~\ref{tab:cv_results} illustrate that the exemplar is capable of executing and evaluating different YOLO model variants confirming that \harmonica can run computer vision pipelines and facilitate comparative studies of adaptation strategies in this setting.

\section{User Study}
\label{sec:user_study}
To assess the usability and utility of \harmonica as an exemplar for sustainable MLOps, we conducted a preliminary user study with practitioners and researchers. The study aimed to evaluate the exemplar's ease of installation, the intuitiveness of the configuration mechanisms, and the clarity of the runtime monitoring dashboards.

\noindent\textbf{5.1 Study Design and Participants}

\noindent We recruited 12 participants with varying levels of experience in AI/ML systems, ranging from 1 to over 8 years. The cohort consisted of students (41.7\%), researchers (41.7\%), and industry practitioners (16.6\%). Participants self-reported their familiarity with Self-Adaptive Systems (SAS) and MLOps on a 5-point Likert scale, resulting in a mean familiarity score of 3.25 ($SD=0.92$).

Participants were tasked with installing \harmonica, deploying one or more adaptation scenarios, as described in Section \ref{sec:empirical_eval}, and interacting with the dashboard. They were asked to modify adaptation policies or thresholds. Following the session, participants were asked to provide feedback via a questionnaire combining Likert scale ratings (1--5) and open-ended questions.

\noindent\textbf{5.2 Quantitative Results:} \noindent The quantitative feedback indicates a high degree of usability. Table~\ref{tab:user_study_results} summarizes the participant ratings.

\begin{table}[h]
\centering
\begin{tabular}{l c c} 
\hline
\textbf{Metric} & \textbf{Mean} & \textbf{Median} \\
\hline
Ease of Installation & 4.58 & 5.0 \\
Intuitiveness of Policy Configuration & 4.08 & 4.0 \\
Clarity of Monitoring Dashboard & 4.41 & 4.5 \\
Likelihood to Recommend & 4.66 & 5.0 \\
\hline
\end{tabular}
\caption{User Study Results (Scale 1--5)}
\label{tab:user_study_results}
\end{table}

\noindent
\textit{Installation and Setup:} Participants found the system easy to deploy, with a mean rating of 4.58/5. This suggests that the exemplar lowers the barrier to entry for experimenting with adaptive MLOps, addressing the reproducibility gap highlighted in Section 1.

\noindent\textit{Observability and Control:} Participants reported a high likelihood of recommending the tool (4.66/5), largely due to its strong runtime observability. The intuitiveness of defining adaptation policies received a slightly lower, though still positive, score of 4.08/5. This suggests that while the mechanism works reliably in practice, the configuration interface would benefit from further refinement.

\noindent\textbf{5.3 Qualitative Feedback}: Participants consistently highlighted the \textit{Live Telemetry}, as shown Figure \ref{fig:dashboard}, and real-time visualization as the exemplar's strongest assets. The ability to view sustainability metric trade-offs in real-time was cited as a key enabler for understanding the underlying adaptation logic. Participants suggested that on-screen tooltips are necessary to explain specific tactics.

\section{Threats to Validity}
\label{sec:threats}

A threat to \textit{Internal Validity} arises from the consistency of results across experimental runs, as energy consumption and performance metrics can be affected by background processes. To mitigate this, all experiments were repeated independently five times, and average results were reported. Additionally, a cooldown period of 20 minutes was enforced between consecutive runs to allow system conditions to stabilize. A threat to \textit{Ecological validity} is that the exemplar runs on a Linux system where energy measurements are obtained using the \texttt{pyRAPL} library, which provides estimates tailored to \texttt{Intel} based hardware. Although absolute energy values may vary when the exemplar is deployed on different processors or measured using alternative tools, we expect the relative trends in adaptation behaviour to remain stable across environments.

\section{Conclusion and Future Works}
\label{sec:conclusion}

This paper presented \harmonica, a reusable exemplar for sustainability aware self-adaptation in MLOps pipelines. By wrapping machine learning workflows in a configurable MAPE-K loop, \harmonica  extends \harmone and enables the observation of runtime trade-offs based on the sustainability goals, and the execution of adaptation tactics without modifying the underlying training or inference code. Through regression and computer vision use cases, we showed that the exemplar can be employed across different ML scenarios, providing a controlled environment for examining adaptation behaviour under varying operating conditions.

Future work will extend the usability and applicability of the exemplar. Although the current implementation employs \texttt{pyRAPL} for energy measurement, our aim is to integrate platform-agnostic energy proxies that generalise across operating systems, virtualised environments, and heterogeneous hardware. This enhancement will make the exemplar more accessible and better aligned with real-world deployment contexts, where consistent and portable energy instrumentation remains an open challenge.

\noindent\textbf{Exemplar Availability:}
The \harmonica is publicly available to support reproducibility and verifiability of the claims presented in this work. We provide a GitHub repository containing the full implementation and documentation, as well as a YouTube video walkthrough that demonstrates installation, setup, and execution of the exemplar. The repository can be accessed at:\\ \textit{\href{https://github.com/sa4s-serc/HarmonE-tool}{https://github.com/sa4s-serc/HarmonE-tool}}. The YouTube video is available at: \textit{\href{https://youtu.be/oiH0wzCPKSs}{https://youtu.be/oiH0wzCPKSs}.  }

\bibliographystyle{ACM-Reference-Format}
\bibliography{references}

\end{document}